\RequirePackage{fix-cm}
\documentclass[smallextended]{svjour3}       
\smartqed  
\usepackage{graphicx}
\begin{document}

\title{Weight hierarchy of a class of linear codes relating to non-degenerate quadratic forms \thanks{This research is supported in part by National Natural Science Foundation of China (Grant Nos.61602342).}
}
\subtitle{}


\author{Fei Li
}


\institute{Fei Li \at
              \email{cczxlf@163.com}           
           \and
Faculty of School of Statistics and Applied Mathematics,
Anhui University of Finance and Economics, Bengbu,  Anhui Province, {\rm 233041}, P.R.China}

\date{Received: date / Accepted: date}

\maketitle

\begin{abstract}
In this paper, we discuss the generalized Hamming weights of a class of linear codes
associated with non-degenerate quadratic forms.
In order to do so, we study the quadratic forms over subspaces of finite field
and obtain some interesting results about subspaces and their dual spaces.
On this basis, we solve all the generalized Hamming weights of these linear codes.

\end{abstract}

\section{Introduction}
\label{intro}
Let $p^{m}$ be a power of an odd prime $p$.
Denote $ F_{p^{m}} $ the finite field with $ p^{m} $
elements and $F_{p^{m}}^{*}$ the multiplicative group of $F_{p^{m}}$.

If $ C $ is a $k$-dimensional $F_{p}$-vector subspace of $F_{p^{n}},$ then it is called
an $[n,k,d]$ $p$-ary linear code with length $ n $ and minimum Hamming distance $d$ \cite{12HP03}.
For linear code $C,$ the concept of generalized Hamming weights(GHW) $ d_{r}(C)(0<r\leq k)$
can be viewed as the extension of Hamming weight (see \cite{15KJ78,20WJ91}). Let
$ [C,r]_{p} $ be the set of all $r$-dimensional $F_{p}$-vector subspaces of $C.$
For $ V \in [C,r]_{p}, $ define $ Supp(V)=\{i|x_{i} \neq 0\ \textrm{for some}\
x = (x_{1}, x_{2}, \cdots , x_{n})\in V\}.$
Then we define the $r$-th generalized Hamming weight(GHW) $ d_{r}(C)$ of linear code $C$ by
$$
d_{r}(C)=\min\{|Supp(V)|V\in [C,r]_{p}\},
$$
In particular, $ d_{1}(C)=d.$ And $ \{d_{i}(C)| 1\leq i \leq k\} $
is called the weight hierarchy of $C. $

Generalized Hamming weight of linear codes has been an interesting topic in both theory
and practice for many years. In 1991, Wei in the paper \cite{20WJ91} presented his classic results,
in which GHW was shown that it can characterise the cryptography performance of linear codes used on the wire-tap
channel of type $II.$
From then on, much more attention was paid to the generalized Hamming weight.
A detailed survey on the results up to 1995 about GHW can be found In \cite{19TV95}.
Afterwards lots of authors devoted themselves to generalized Hamming weight about particular classes of codes \cite{1BL14,2BM00,3CC97,7DF14,11HP98,13JF17,14JL97,21XL16,22YL15}.
Recently, Minghui Yang et al. in their work \cite{22YL15}
gave a very constructive method for the GHWs of irreducible cyclic codes.
Generally, it is hard to settle the weight hierarchy of a linear code.

Ding et al. proposed a generic construction of linear code as below (\cite{5DJ15,6DD14}).
Let $Tr$ be the trace function from $F_{p^{m}}$ to $F_p$
and $ D= \{d_{1},d_{2},\cdots,d_{n}\}$ be contained in $F_{p^{m}}^{*}. $ Define
a $p$-ary linear code $ C_{D} $ with length $ n $ as following.
\begin{eqnarray}\label{defcode1}
         C_{D}=\{\left( Tr(xd_1), Tr(xd_2),\ldots, Tr(xd_{n})\right):x\in F_{p^{m}}\}
\end{eqnarray}
and $ D $ is called the defining set. Many classes of linear codes with few weights
were obtained by choosing properly defining sets \cite{9DL16,17LY14,23YY17,24ZL16,26TX17}.
In this paper, we discuss the generalized Hamming weights of a class of
$p$-ary linear codes $ C_{D}$, whose defining set was chosen to be
\begin{eqnarray}\label{defcode2}
         D=D_{a}=\{x\in F_{p^{m}}^{*}|f(x)=a\}, \ a \in F_{p}.
\end{eqnarray}
Here $f$ is a non-degenerate quadratic form over $F_{p^{m}}.$

And the weight hierarchy of $C_{D_{0}}$ can be deduced by Theorem 9 in \cite{27WZ94},
in which Mr. Wan used the theory of finite projective geometry for the purpose.

In reference \cite{25DD15}, Kelan Ding and Cunsheng Ding presented the linear
codes $C_{D_{a}}$ of the case $a=0$ for the special quadratic form $Tr(X^{2})$ 
and determined their weight distributions.
In \cite{24ZL16} and \cite{8DW17}, the weight distributions of $C_{D_{a}}$ 
for general quadratic forms were settled.
By the main results of them, we know that $C_{D_{a}}$ is an $m$-dimensional linear code.
So we can employ a general formula for linear codes defined in (1) to calculate $ d_{r}(C_{D_{a}}).$
We give it in the following lemma.

\par \vskip 0.2 cm
{\bf Lemma 1.(Theorem 1, \cite{18LF17})}\
For each $ r (1\leq r \leq m), $  if the dimension of $ C_{D} $
is $m,$ then $d_{r}(C_{D})= n-\max\{|D \bigcap H|: H \in [F_{p^{m}},m-r]_{p}\}. $

\par \vskip 0.2 cm
The rest of this paper is organised as follows: in Section 2, we present some basic definitions and results
on quadratic form over subspace of finite fields and dual spaces; in Section 3,
using the results in Section 2, we give all the generalized Hamming weights of linear codes
defined in (2); in Section 4, we give the conclusion of this paper.

\par \vskip 0.2 cm
\section{Quadratic Form over Subspace of Finite Fields}

We may view $F_{p^{m}}$ as an $m$-dimensional vector space over $F_{p}.$
Namely, $F_{p^{m}}$ is $F_{p}$-linearly isomorphic to $F_{p}^{m}.$ We
fix $m$ elements $\upsilon_{1},\upsilon_{2},\cdots,\upsilon_{m} \in F_{p^{m}}$
as a basis and express each vector $X\in F_{p^{m}}$ in the unique form
$X=x_{1}\upsilon_{1}+x_{2}\upsilon_{2}+x_{m}\upsilon_{m}, $ for some
$x_{1},x_{2},\cdots,x_{m}\in F_{p}.$ By the isomorphism, we can write
$X=(x_{1},x_{2},\cdots,x_{m})^{T}, $ where $T$ represents the transpose of a matrix.

Now we fix a basis $\upsilon_{1},\upsilon_{2},\cdots,\upsilon_{m} \in F_{p^{m}}.$
Recall that a map $ f: F_{p^{m}} \rightarrow F_{p}$ is called a quadratic form 
over $F_{p^{m}}$ with values in $F_{p}$ if 
$$
f(X)=f(x_{1},x_{2},\cdots,x_{m})=\sum_{1\leq i,j\leq m}a_{ij}x_{i}x_{j}, a_{ij}\in F_{p},
$$
for each element $X=x_{1}\upsilon_{1}+x_{2}\upsilon_{2}+x_{m}\upsilon_{m}.$
Denote $ F(X,Y)=\frac{1}{2}[f(X+Y)-f(X)-f(Y)],$ so $f(X)=F(X,X).$
If $a_{ij}=a_{ji}, $ we can write $ f(X)=X^{T}AX, $ where $A$ is the
symmetric matrix $(a_{ij})_{1\leq i,j\leq m}.$ The rank $R_{f}$ of quadratic form $f$
is defined to be the rank of matrix $A.$ It is well-known that $a_{ii}=f(\upsilon_{i}),
a_{ij}=F(\upsilon_{i},\upsilon_{j}).$

For a subspace $H\subseteq F_{p^{m}}, $ the dual space $H^{\bot}$ of $H$ under
quadratic form $f$ is defined by 
$$
H^{\bot}=\{x\in F_{p^{m}}|f(x+y)=f(x)+f(y)\ \textrm{for each} \ y \in H\}.
$$
Let $G$ be another subspace of $F_{p^{m}}.$ If $ f(x+y)=f(x)+f(y) $ for each $x\in H$ and $y\in G,$ we say that
$H$ and $G$ are perpendicular to each other under quadratic form $f.$ It is denoted by $H\perp G.$
It is well-known that $ R_{f}+\dim(F_{p^{m}}^{\bot})=m.$ If $R_{f}=m, $ we say that the quadratic form $f$
is non-degenerate. And the discriminant $\Delta(f)$ of the quadratic form $f$ is defined as the determinant
$\det(A)$ of $A.$ We can find a invertible matrix $M$ such that $M^{T}AM$ is a diagonal matrix
$\Lambda=diag(\lambda_{1},\lambda_{2},\cdots,\lambda_{R_{f}},0,\cdots,0).$ When $0<R_{f}<m, $ we
also can define the discriminant $\Delta(f)=\lambda_{1}\cdot\lambda_{2}\cdots\lambda_{R_{f}}.$
When $R_{f}=0, $ we define $\Delta(f)=1.$
It is easy to see that $\Delta(f)$ varies with the change of the basis of $F_{p^{m}}.$
But $\overline{\eta}(\Delta(f))$ is an invariant, which is called the sign of
the quadratic form $f.$ We denote it by $\epsilon_{f}.$ Here $\overline{\eta}$ is the quadratic
character of $F_p.$ Meanwhile, $\overline{\eta}(0)$ is defined to be zero.

Let $H$ be a $d$-dimensional subspace of $F_{p^{m}}.$ Restricting the quadratic form
$f$ on $H,$ we get a quadratic form over $H$ in $d$ variables. It is denoted by $f|_{H}.$
Similarly, we define the dual space $H^{\bot}_{f|_{H}}$ of $H$ under $f|_{H}$ in itself by
$$
H^{\bot}_{f|_{H}}=\{x\in H|f(x+y)=f(x)+f(y)\ \textrm{for each} \ y \in H\}.
$$ 
Let $R_{f|_{H}}, \Delta(f|_{H})$ be the rank and discriminant
of $f|_{H}$ over $H,$ respectively. Obviously, $H^{\bot}_{f|_{H}}=H\bigcap H^{\bot}$ and $R_{f|_{H}}=d-\dim(H^{\bot}_{f|_{H}}).$

For $k$ elements
$\beta_{1},\beta_{2},\cdots, \beta_{k}\in F_{p^{m}},$ the discriminant
$\Delta(\beta_{1},\beta_{2},\cdots, \beta_{k})$ of them is defined as the determinant of
the matrix $M=(F(\beta_{i},\beta_{j}))_{1\leq i,j\leq k}.$

From now on, we discuss only the non-degenerate quadratic forms. Its
symmetric matrix is $ (F(\upsilon_{i},\upsilon_{j}))_{1\leq i,j\leq m} $ if we
fix an $F_{p}$-basis $\upsilon_{1},\upsilon_{2},\cdots,\upsilon_{m}$ of $ F_{p^{m}}. $
Thus we have that the discriminant $\Delta(f)=\Delta(\upsilon_{1},\upsilon_{2}\cdots \upsilon_{m}).$
For a subspace $H$ of $F_{p^{m}}, $
we know that $\dim(H)+\dim(H^{\bot})=m $ and denote $R_{H}=R_{f|_{H}},
\Delta_{H}=\Delta(f|_{H}).$ If we fix an $F_{p}$-basis $ \mu_{1},\mu_{2}\cdots \mu_{d}$ of $H,$
then $\Delta_{H}=\Delta(\mu_{1},\mu_{2}\cdots \mu_{d}).$
Let $v(x)$ be a function over $F_{p}$ defined by $v(x)=p-1$ if $x=0,$ otherwise $v(x)=-1.$
For $a\in F_{p},$ set $\overline{D}_{a}=\{x\in F_{p^{m}}|f(x)=a\}.$ Then we have following results.

\par  \vskip 0.5 cm
{\bf Proposition 1.}\
Let $H$ be a $d$-dimensional($d>0$) subspace of $F_{p^{m}},$ then
$$
|H\bigcap \overline{D}_{a}|=\left\{\begin{array}{ll}
p^{d-1}+v(a)\overline{\eta}((-1)^{\frac{R_{H}}{2}}\Delta_{H})p^{d-\frac{R_{H}+2}{2}}, & \textrm{if\ } \ R_{H}\equiv0(\textrm{mod}2), \\
p^{d-1}+\overline{\eta}((-1)^{\frac{R_{H}-1}{2}}a\Delta_{H})p^{d-\frac{R_{H}+1}{2}}, & \textrm{if\ } \ R_{H}\equiv1(\textrm{mod}2).
\end{array}
\right.
$$

{\bf Proof. } \ It is the natural result of quadratic forms from
Theorem 6.26 and 6.27 in \cite{16LN97}. We omit the details.

\par \vskip 0.5 cm
{\bf Proposition 2. } \ For each $ r(0<2r < m), $  there exist an $r$-dimensional subspace
$H\subseteq F_{p^{m}}(m>2)$ such that $H\subseteq H^{\bot},$ i.e., $H\perp H.$

{\bf Proof. } \
It suffices to prove there exist $r$ elements $\alpha_{1},\alpha_{2},\ldots,\alpha_{r}\in F_{p^{m}}$
satisfying the following two conditions:
\begin{enumerate}
\item $F(\alpha_{i},\alpha_{j})=0, 1\leq i,j\leq r $;
\item $\alpha_{1},\alpha_{2},\ldots,\alpha_{r}$ are linearly independent over $F_{p}$.
\end{enumerate}
If $r=1,$ by Lemma 1, $ \overline{D}_{0}\neq\{0\}.$ So it is true for $r=1.$
Suppose we have $k$ elements $ \alpha_{1},\alpha_{2},\ldots,\alpha_{k},$
which satisfy the two conditions. It might as well assume $2k+2< m.$ Otherwise, the proof is
finished. Let $ H_{k}=\langle \alpha_{1},\alpha_{2},\ldots,\alpha_{k}\rangle.$
So $ H^{\bot}_{k}$ is an $(m-k)$-dimension subspace.
By Proposition 1, $|\overline{D}_{0}\bigcap H^{\bot}_{k}|=p^{m-k-1}\pm(p-1)p^{m-k-1-\frac{t}{2}}$
if $t$ is even, otherwise $|\overline{D}_{0}\bigcap H^{\bot}_{k}|=p^{m-k-1}.$ Here $t=R_{H^{\bot}_{k}}.$
Let $ V_{k}=\{x\in H^{\bot}_{k}|F(x,y)=0, y \in H^{\bot}_{k}\}.$
So $ \dim(V_{k})=m-k-t.$ Since $ V_{k}\subset H_{k}, m-k-t\leq k. $ Hence $t\geq m-2k>2.$
Therefore $|\overline{D}_{0}\bigcap H^{\bot}_{k}|>p^{m-k-1}-(p-1)p^{m-k-2}=p^{m-k-2}>p^{k}.$ We obtain that there
is an element $ \alpha_{k+1}\in (\overline{D}_{0}\bigcap H^{\bot}_{k})\backslash H_{k}.$ So
$ \alpha_{1},\alpha_{2},\ldots,\alpha_{k}, \alpha_{k+1}$ are $(k+1)$ elements that match the conditions
as above. By mathematical induction, we finish the proof.

\par \vskip 0.5 cm
{\bf Proposition 3. } \ Let $ m=2s>2.$ There exists an $s$-dimensional subspace $H_{s}\subset F_{p^{m}}$
such that $H_{s}=H^{\bot}_{s}$ if and only if $\epsilon_{f}=(-1)^{\frac{m(p-1)}{4}}.$

{\bf Proof. } \ By Proposition 2, there exist $(s-1)$ elements
$ \alpha_{1},\alpha_{2},\ldots, \alpha_{s-1}$ that match the following conditions:
\begin{enumerate}
\item $F(\alpha_{i},\alpha_{j})=0, 1\leq i,j\leq s-1 $;
\item $\alpha_{1},\alpha_{2},\ldots,\alpha_{s-1}$ are linearly independent over $F_{p}$.
\end{enumerate}
Let
$
H_{s-1}=\langle \alpha_{1},\alpha_{2},\ldots,\alpha_{s-1}\rangle.
$
So $ H^{\bot}_{s-1}$ is an $(s+1)$-dimension subspace and the rank of quadratic form
$ f(x)$ over $ H^{\bot}_{s-1}$ is $2.$
Let
$$
H^{\bot}_{s-1}=\langle \alpha_{1},\alpha_{2},\ldots,\alpha_{s-1},\alpha_{s},\alpha_{s+1}\rangle
$$
and $ H_{2}=\langle \alpha_{s},\alpha_{s+1}\rangle.$ It is easy to see
$\Delta_{H^{\bot}_{s-1}}=\Delta(\alpha_{s},\alpha_{s+1}) $ and $H_{2}\bigcap H^{\bot}_{2}=\{0\}.$ Suppose
$$
 H^{\bot}_{2}=\langle \alpha_{1},\alpha_{2},\ldots,\alpha_{s-1},\beta_{1},\beta_{2},\ldots,\beta_{s-1}\rangle
$$
So $\Delta_{m}=\Delta(\alpha_{1},\ldots,\alpha_{s-1},\alpha_{s},\alpha_{s+1},
\beta_{1},\ldots,\beta_{s-1})=(-1)^{s+1}\det(M^{2})\Delta_{H^{\bot}_{s-1}}.$
Here $M$ is the square matrix $(F(\alpha_{i},\beta_{j}))_{1\leq i,j\leq s-1}.$
By Proposition 1, we have $|H^{\bot}_{s-1}\bigcap \overline{D}_{0}|=p^{s}+\overline{\eta}((-1)^{s}\Delta(f))(p-1)p^{s-1}.$
If $\overline{\eta}(\Delta(f))=(-1)^{\frac{m(p-1)}{4}},$ then $|H^{\bot}_{s-1}\bigcap \overline{D}_{0}|=2p^{s}-p^{s-1}>p^{s-1}. $
So we can choose one element $ \alpha_{s}\in (H^{\bot}_{s-1}\bigcap \overline{D}_{0})\backslash H_{s-1}$ and set
$H_{s}=\langle \alpha_{1},\alpha_{2},\ldots,\alpha_{s-1},\alpha_{s}\rangle.$ It is easy to check
$ H^{\bot}_{s}=H_{s}.$ If $\overline{\eta}(\Delta(f))=-(-1)^{\frac{m(p-1)}{4}},$ then $|H^{\bot}_{s-1}\bigcap \overline{D}_{0}|=p^{s-1}. $
So $H^{\bot}_{s-1}\bigcap \overline{D}_{0}$ must be $H_{s-1}.$ In this case, the conclusion of the proposition
is untenable. We complete the proof.

\section{Weight Hierarchy of Linear Codes of (2)}

\par \vskip 0.2 cm
In this section, we settle the weight hierarchy of the linear code defined in (2).

\subsection{\bf Case 1:\ $a\neq0$}

\par  \vskip 0.2 cm
{\bf Theorem 1.}\
If $m$ is even, then for the linear codes defined in (2) we have
$$
d_{r}(C_{D_{a}})=\left\{\begin{array}{ll}
p^{m-1}-p^{m-r-1}-((-1)^{\frac{m(p-1)}{4}}\epsilon_{f}+1)p^{\frac{m-2}{2}}, & \textrm{if\ } \ 1\leq r\leq \frac{m}{2}, \\
p^{m-1}-2p^{m-r-1}-(-1)^{\frac{m(p-1)}{4}}\epsilon_{f}p^{\frac{m-2}{2}}, & \textrm{if\ } \ \frac{m}{2}\leq r< m, \\
p^{m-1}-(-1)^{\frac{m(p-1)}{4}}\epsilon_{f}p^{\frac{m-2}{2}}, & \textrm{if\ } \  r= m.
\end{array}
\right.
$$

\par \vskip 0.2 cm
{\bf Proof. } \ We only give the proof of the case: $\epsilon_{f}=-(-1)^{\frac{m(p-1)}{4}}.$
The proof of the other case: $\epsilon_{f}=(-1)^{\frac{m(p-1)}{4}}$ is similar.

For $1\leq r< \frac{m}{2}, $ let $t=R_{H_{m-r}},$ the rank of $ f(x)$ over
some $(m-r)$-dimensional subspace $ H_{m-r}.$
By Proposition 2,3, $ 0\leq \dim(H_{m-r}\bigcap H^{\bot}_{m-r})\leq r.$
So $0\leq m-r-t\leq r. $ Hence $ m-2r\leq t\leq m-r.$

If $t=m-2r,$ we may construct
$$ H_{m-r}=\langle \alpha_{1},\alpha_{2},\ldots,\alpha_{m-2r},\beta_{1},\beta_{2},\ldots,\beta_{r}\rangle,
 H_{r}=H^{\bot}_{m-r}=\langle \beta_{1},\beta_{2},\ldots,\beta_{r}\rangle,
$$
$$ H_{m-2r}=\langle \alpha_{1},\alpha_{2},\ldots,\alpha_{m-2r}\rangle,
 H_{2r}=H^{\bot}_{m-2r}=\langle \beta_{1},\beta_{2},\ldots,\beta_{r},\beta_{r+1},\ldots,\beta_{2r}\rangle.
$$
Because $ H_{m-2r}\bigcap H^{\bot}_{m-2r}=\{0\}, $
$$
\Delta_{m}=\Delta(\alpha_{1},\alpha_{2},\ldots,\alpha_{m-2r}
,\beta_{1},\beta_{2},\ldots,\beta_{r},\beta_{r+1},\ldots,\beta_{2r})
$$
$$
=\Delta_{H_{m-r}}\Delta(\beta_{1},\beta_{2},\ldots,
\beta_{r},\beta_{r+1},\ldots,\beta_{2r})=(-1)^{r}\Delta_{H_{m-r}}\det(M^{2}).
$$
Here $M$ is the square matrix $M=(F(\beta_{i},\beta_{j}))_{1\leq i\leq r,r+1\leq j\leq 2r}.$
Hence for $t= m-2r, $ by Proposition 1,
$|D_{a}\bigcap H_{m-r}|=p^{m-r-1}-\overline{\eta}((-1)^{\frac{m}{2}})\epsilon_{f}p^{\frac{m-2}{2}}
=p^{m-r-1}+p^{\frac{m-2}{2}}. $ So by Proposition 1, we have that
$\max\{|D_{a} \bigcap H|: H \in [F_{p^{m}},m-r]_{p}\}=p^{m-r-1}+p^{\frac{m-2}{2}}.$

For $\frac{m}{2}\leq r<m,$ by Proposition 2, there is such an $(m-r-1)$-dimensional subspace
$H_{m-r-1}=\langle \alpha_{1},\alpha_{2},\ldots,\alpha_{m-r-1}\rangle$
that $H_{m-r-1}\subseteq H_{m-r-1}^{\bot}.$ By Proposition 1,
$|D_{b}\bigcap H^{\bot}_{m-r-1}|>0$ with $b\in F_{p}^{*}.$
We choose an element $\alpha_{m-r}\in(D_{b}\bigcap H^{\bot}_{m-r-1})$ and get
an $(m-r)$-dimensional subspace
$$
H_{m-r}=\langle \alpha_{1},\alpha_{2},\ldots,\alpha_{m-r-1},\alpha_{m-r}\rangle.
$$
By the construction of $H_{m-r},$ we know that the rank of $f(x)$ over $H_{m-r}$ is $1$
and $\Delta_{H_{m-r}}=b.$ So by Proposition 1, $|D_{a}\bigcap H_{m-r}|=p^{m-r-1}+\overline{\eta}(ab)p^{m-r-1}.$
Also by Proposition 1, we have that
$\max\{|D_{a} \bigcap H|: H \in [F_{p^{m}},m-r]_{p}\}=2p^{m-r-1}.$
By Proposition 1, we have $|D_{a}|=p^{m-1}-\epsilon_{f}(-1)^{\frac{m(p-1)}{4}}p^{\frac{m-2}{2}}.$
Then the desired results follow directly from Lemma 1. And we complete the proof.

\par  \vskip 0.2 cm
{\bf Theorem 2.}\
If $m$ is odd and $\overline{\eta}(a)=(-1)^{\frac{(m-1)(p-1)}{4}}\epsilon_{f},$ then for the linear codes defined in (2) we have
$$
d_{r}(C_{D_{a}})=\left\{\begin{array}{ll}
p^{m-1}-p^{m-r-1}, & \textrm{if\ } \ 1\leq r< \frac{m}{2}, \\
p^{m-1}+p^{\frac{m-1}{2}}-2p^{m-r-1}, & \textrm{if\ } \ \frac{m}{2}< r< m, \\
p^{m-1}+p^{\frac{m-1}{2}}, & \textrm{if\ } \ r= m.
\end{array}
\right.
$$

\par \vskip 0.2 cm
{\bf Proof. } \
For $1\leq r< \frac{m}{2}, $ we may construct subspaces
$$ H_{m-r}=\langle \alpha_{1},\alpha_{2},\ldots,\alpha_{m-2r},\beta_{1},\beta_{2},\ldots,\beta_{r}\rangle,
 H_{r}=H^{\bot}_{m-r}=\langle \beta_{1},\beta_{2},\ldots,\beta_{r}\rangle,
$$
$$ H_{m-2r}=\langle \alpha_{1},\alpha_{2},\ldots,\alpha_{m-2r}\rangle,
 H_{2r}=H^{\bot}_{m-2r}=\langle \beta_{1},\beta_{2},\ldots,\beta_{r},\beta_{r+1},\ldots,\beta_{2r}\rangle.
$$
Because $ H_{m-2r}\bigcap H^{\bot}_{m-2r}=\{0\}, $
$$
\Delta_{m}=\Delta(\alpha_{1},\alpha_{2},\ldots,\alpha_{m-2r}
,\beta_{1},\beta_{2},\ldots,\beta_{r},\beta_{r+1},\ldots,\beta_{2r})
$$
$$
=\Delta_{H_{m-r}}\Delta(\beta_{1},\beta_{2},\ldots,
\beta_{r},\beta_{r+1},\ldots,\beta_{2r})=(-1)^{r}\Delta_{H_{m-r}}\det(M^{2}).
$$
Here $M$ is the square matrix $M=(F(\beta_{i},\beta_{j}))_{1\leq i\leq r,r+1\leq j\leq 2r}.$
Note that $\overline{\eta}(a)=(-1)^{\frac{(m-1)(p-1)}{4}}\epsilon_{f}.$ So
$|D_{a}\bigcap H_{m-r}|=p^{m-r-1}+\overline{\eta}((-1)^{\frac{m}{2}}a)\epsilon_{f}p^{\frac{m-1}{2}}
=p^{m-r-1}+p^{\frac{m-1}{2}}. $ By Proposition 1, we have that
$\max\{|D_{a} \bigcap H|: H \in [F_{p^{m}},m-r]_{p}\}=p^{m-r-1}+p^{\frac{m-1}{2}}.$

For the case: $\frac{m}{2}< r< m,$ the proof is the same as that of Theorem 1.
By Proposition 1, we have
$|D_{a}|=p^{m-1}+\epsilon_{f}(-1)^{\frac{m(p-1)}{4}}\overline{\eta}(a)p^{\frac{m-1}{2}}.$
Then the desired conclusions follow from Lemma 1. And the proof is completed.

\par  \vskip 0.2 cm
{\bf Theorem 3.}\
If $m$ is odd and $\overline{\eta}(a)=-(-1)^{\frac{(m-1)(p-1)}{4}}\epsilon_{f},$
then for the linear codes defined in (2) we have
$$
d_{r}(C_{D_{a}})=\left\{\begin{array}{ll}
p^{m-1}-p^{m-r-1}-p^{\frac{m-1}{2}}-p^{\frac{m-3}{2}}, & \textrm{if\ } \ 1\leq r< \frac{m}{2}, \\
p^{m-1}-p^{\frac{m-1}{2}}-2p^{m-r-1}, & \textrm{if\ } \ \frac{m}{2}< r< m, \\
p^{m-1}-p^{\frac{m-1}{2}}, & \textrm{if\ } \ r= m.
\end{array}
\right.
$$

\par \vskip 0.2 cm
{\bf Proof. } \
The proof is similar to those of Theorem 1 and Theorem 2. We omit the details.

\subsection{\bf Case 2:\ $a=0$}

\par  \vskip 0.2 cm
{\bf Theorem 4.}\
If $m$ is even and $\epsilon_{f}=(-1)^{\frac{m(p-1)}{4}},$ then for the linear codes defined in (2) we have
$$
d_{r}(C_{D_{0}})=\left\{\begin{array}{ll}
p^{m-1}-p^{m-r-1}, & \textrm{if\ } \ 1\leq r\leq \frac{m}{2}, \\
p^{m-1}+(p-1)p^{\frac{m-2}{2}}-p^{m-r}, & \textrm{if\ } \ \frac{m}{2}\leq r\leq m.
\end{array}
\right.
$$

\par \vskip 0.2 cm
{\bf Proof. } \ Denote by $H_{k}$ a $k$-dimensional subspace in $F_{p^{m}}.$
By Lemma 1 and Proposition 2 and 3 as above, it is easy to know $ d_{r}(C_{D_{0}})=|\overline{D}_{0}|-p^{m-r}, $
if $\frac{m}{2}\leq r\leq m.$

For $1\leq r< \frac{m}{2}, $ let $t=R_{H_{m-r}},$ the rank of quadratic form $ f(x)$ over
some $(m-r)$-dimensional subspace $ H_{m-r}.$
For $ 0\leq \dim(H_{m-r}\bigcap H^{\bot}_{m-r})\leq r,$
so $0\leq m-r-t\leq r. $ Hence $ m-2r\leq t\leq m-r.$

If $r=1,$ then $d_{r}(C_{D_{0}})$ is the minimum Hamming distance of $C_{D_{0}}.$ The result for this case
can be obtained in reference \cite{24ZL16}.

For the case $\frac{m}{2}> r\geq2,$ if $t=m-2r,$ we may construct
$$ H_{m-r}=\langle \alpha_{1},\alpha_{2},\ldots,\alpha_{m-2r},\beta_{1},\beta_{2},\ldots,\beta_{r}\rangle,
 H_{r}=H^{\bot}_{m-r}=\langle \beta_{1},\beta_{2},\ldots,\beta_{r}\rangle,
$$
$$ H_{m-2r}=\langle \alpha_{1},\alpha_{2},\ldots,\alpha_{m-2r}\rangle,
 H_{2r}=H^{\bot}_{m-2r}=\langle \beta_{1},\beta_{2},\ldots,\beta_{r},\beta_{r+1},\ldots,\beta_{2r}\rangle.
$$
Because $ H_{m-2r}\bigcap H^{\bot}_{m-2r}=\{0\}, $
$$
\Delta_{m}=\Delta(\alpha_{1},\alpha_{2},\ldots,\alpha_{m-2r}
,\beta_{1},\beta_{2},\ldots,\beta_{r},\beta_{r+1},\ldots,\beta_{2r})
$$
$$
=\Delta_{H_{m-r}}\Delta(\beta_{1},\beta_{2},\ldots,
\beta_{r},\beta_{r+1},\ldots,\beta_{2r})=(-1)^{r}\Delta_{H_{m-r}}\det(M^{2}).
$$
Here $M$ is the square matrix $M=(F(\beta_{i},\beta_{j}))_{1\leq i\leq r,r+1\leq j\leq 2r}.$
Notice that $\epsilon_{f}=(-1)^{\frac{m(p-1)}{4}}.$
Hence for $t= m-2r, $ by Proposition 1, $|\overline{D}_{0}\bigcap H_{m-r}|=p^{m-r-1}+\overline{\eta}((-1)^{\frac{m-2r}{2}}\Delta_{H_{m-r}})(p-1)p^{m-r-1-\frac{m-2r}{2}}
=p^{m-r-1}+(p-1)p^{\frac{m-2}{2}}. $ So by Proposition 1, we have that
$\max\{|\overline{D}_{0} \bigcap H|: H \in [F_{p^{m}},m-r]_{p}\}=p^{m-r-1}+(p-1)p^{\frac{m-2}{2}}.$
Then the desired results follow directly from Lemma 1. And we complete the proof.

\par  \vskip 0.2 cm
{\bf Theorem 5.}\
If $m=2s>2$ is even and $\epsilon_{f}=-(-1)^{\frac{m(p-1)}{4}},$ then for the linear codes defined in (2) we have
$$
d_{r}(C_{D_{0}})=\left\{\begin{array}{ll}
(p-1)(p^{m-2}-p^{\frac{m-2}{2}}), & \textrm{if\ } \ r= 1, \\
p^{m-1}-p^{m-r-1}-(p-1)(p^{\frac{m-2}{2}}+p^{\frac{m-4}{2}}), & \textrm{if\ } \ 2\leq r\leq \frac{m}{2}, \\
p^{m-1}-p^{m-r}-(p-1)p^{\frac{m-2}{2}}, & \textrm{if\ } \ \frac{m}{2}< r\leq m.
\end{array}
\right.
$$

\par \vskip 0.2 cm
{\bf Proof. } \
We only give the proof of the two cases: $2\leq r< \frac{m}{2}$
and $r=\frac{m}{2}.$ The proofs of the remaining cases are the same as that of Theorem 1.

For $2\leq r< \frac{m}{2},$ let the symbols be as in the proof of Theorem 1.
By the proof of Theorem 1, we have
$|\overline{D}_{0}\bigcap H_{m-r}|=p^{m-r-1}-(p-1)p^{\frac{m-2}{2}}$ if $t=m-2r.$

If $t=m-2r+2,$ by Proposition 3, we can choose such a subspace
$ G_{r-2}=\langle \varepsilon_{1},\varepsilon_{2},\ldots,\varepsilon_{r-2}\rangle $
that $ G_{r-2}\subset G^{\bot}_{r-2}.$ So the rank of $f(x)$
over $G^{\bot}_{r-2}$ is $m-2r+4.$ By Proposition 1,
$|\overline{D}_{a}\bigcap G^{\bot}_{r-2}|=p^{m-r+2-1}\pm p^{m-r+2-1-\frac{m-2r+4}{2}}>0$ with $ a\in F^{*}_{p}.$
So we can choose an nonzero element $\varepsilon_{r-1}\in \overline{D}_{a}\bigcap G^{\bot}_{r-2}$
and get an $(r-1)$-dimensional subspace
$ G_{r-1}=\langle \varepsilon_{1},\varepsilon_{2},\ldots,\varepsilon_{r-2},\varepsilon_{r-1}\rangle.$
It is easy to see that the rank of $f(x)$ over $G_{r-1}$ is $1$
and $ G_{r-1}\bigcap G^{\bot}_{r-1}=G_{r-2}.$
So the rank of $f(x)$ over $G^{\bot}_{r-1}$ is $m-2r+3.$
By Proposition 1,
$|\overline{D}_{b}\bigcap G^{\bot}_{r-1}|=p^{m-r}\pm p^{m-r+1-\frac{m-2r+4}{2}}>p^{r-1}$ with $ b\in F^{*}_{p}.$
So we obtain an $r$-dimensional subspace $ H_{r}=\langle \varepsilon_{1},\varepsilon_{2},\ldots,\varepsilon_{r-1},\varepsilon_{r}\rangle$
by choosing $ \varepsilon_{r}\in(\overline{D}_{b}\bigcap G^{\bot}_{r-1})\setminus G_{r-1}$
and the rank of $f(x)$ over $H_{r}$ is $2.$ So we can construct another $(m-r)$-dimensional subspace
$ H_{m-r}=H^{\bot}_{r}$ and set
$$
 H_{m-r}=
\langle \gamma_{1},\gamma_{2},\ldots,\gamma_{m-2r+2},\varepsilon_{1},\varepsilon_{2},\ldots,\varepsilon_{r-2}\rangle.
$$
Let $H_{m-2r+2}=\langle \gamma_{1},\gamma_{2},\ldots,\gamma_{m-2r+2}\rangle$ and set
$$
H^{\bot}_{m-2r+2}=\langle \varepsilon_{1},\ldots,\varepsilon_{r-1},\varepsilon_{r},\varepsilon_{r+1},\ldots,\varepsilon_{2r-2}\rangle.
$$
Because $R_{H_{m-r}}=m-2r+2,$ we get $H^{\bot}_{m-2r+2}\bigcap H_{m-2r+2}=\{0\}$ and
$$
\Delta_{m}=\Delta(\gamma_{1},\gamma_{2},\ldots,\gamma_{m-2r+2},
\varepsilon_{1},\ldots,\varepsilon_{r-1},\varepsilon_{r},\varepsilon_{r+1},\ldots,\varepsilon_{2r-2})
$$
$$
=\Delta_{H_{m-r}}\cdot\Delta(\varepsilon_{1},\ldots,\varepsilon_{r-1},\varepsilon_{r},
\varepsilon_{r+1},\ldots,\varepsilon_{2r-2})
$$
$$
=(-1)^{r}\Delta_{H_{m-r}}ab\det(M^{2}).
$$
Here $M$ is the square matrix $M=(F(\varepsilon_{i},\varepsilon_{j}))_{1\leq i\leq r-2,r+1\leq j\leq 2r-2}.$
According to the selection of $\varepsilon_{r-1},\varepsilon_{r},$ we know that $ \overline{\eta}(\Delta_{H_{m-r}})$
may take $1$ or $-1.$
Hence for $t= m-2r+2, $ by Proposition 1,
$|\overline{D}_{0}\bigcap H_{m-r}|=p^{m-r-1}\pm(p-1)p^{\frac{m}{2}-2}. $

For $r=\frac{m}{2},$ let $H_{s}$ be an $s$-dimensional subspace.
We have $1\leq R_{H_{s}}\leq s$ since $\epsilon_{f}=-(-1)^{\frac{m(p-1)}{4}}$
by Proposition 3.
If $R_{H_{s}}=1,$ then $|\overline{D}_{0}\bigcap H_{s}|=p^{s-1}.$ If $R_{H_{s}}=2,$
then $|\overline{D}_{0}\bigcap H_{s}|=p^{s-1}+ \overline{\eta}(-\Delta_{H_{s}})(p-1)p^{s-2}.$
By Proposition 2 and Proposition 3, there exists an $(s-2)$-dimensional
subspace $H_{s-2}$ such that $H_{s-2}\subset H^{\bot}_{s-2}.$ Set
$
H_{s-2}=\langle \alpha_{1},\alpha_{2},\ldots,\alpha_{s-2}\rangle.
$
So the rank of quadratic form $f$ over $H^{\bot}_{s-2}$ is $4.$ By Proposition 1,
$|\overline{D}_{a}\bigcap H^{\bot}_{s-2}|=p^{s+1}\pm p^{s-1}>0$ with $a\in F_{p}^{*}.$ We
choose an element $\alpha_{s-1}\in \overline{D}_{a}\bigcap H^{\bot}_{s-2}$ and get
an $(s-1)$-dimensional subspace
$H_{s-1}=\langle \alpha_{1},\alpha_{2},\ldots,\alpha_{s-2},\alpha_{s-1}\rangle.$
The rank of quadratic form $f$ over $H^{\bot}_{s-1}$ is $3.$
Also by Proposition 1, $|\overline{D}_{b}\bigcap H^{\bot}_{s-1}|=p^{s}\pm p^{s-1}>p^{s-1}$ with $b\in F_{p}^{*}.$
So we choose an element $\alpha_{s}\in(\overline{D}_{b}\bigcap H^{\bot}_{s-1})\setminus H_{s-1}$ and get
an $s$-dimensional subspace
$H_{s}=\langle \alpha_{1},\alpha_{2},\ldots,\alpha_{s-2},\alpha_{s-1},\alpha_{s}\rangle.$
By the construction of $H_{s},$ we know that the rank of $f$ over $H_{s}$ is $2$
and $\overline{\eta}(\Delta_{H_{s}})=\overline{\eta}(\Delta(\alpha_{s-1},\alpha_{s}))$ may take $-1$ or $1.$
So by Proposition 1, $|\overline{D}_{0}\bigcap H_{s}|=p^{s-1}\pm (p-1)p^{s-2}.$

In a word, we have that when $2\leq r\leq \frac{m}{2},$
$\max\{|\overline{D}_{0} \bigcap H|: H \in [F_{p^{m}},m-r]_{p}\}=p^{m-r-1}+(p-1)p^{\frac{m-4}{2}}.$
Then we arrive at the conclusion by Lemma 1. And the proof is finished.

\par  \vskip 0.2 cm
{\bf Theorem 6.}\
If $m$ is odd, then for the linear codes defined in (2) we have
$$
d_{r}(C_{D_{0}})=\left\{\begin{array}{ll}
p^{m-1}-p^{m-r-1}-(p-1)p^{\frac{m-3}{2}}, & \textrm{if\ } \ 1\leq r< \frac{m}{2}, \\
p^{m-1}-p^{m-r}, & \textrm{if\ } \ \frac{m}{2}< r\leq m.
\end{array}
\right.
$$

\par \vskip 0.2 cm
{\bf Proof. } \
We only give the proof of the case: $2\leq r< \frac{m}{2}.$
For an $(m-r)$-dimensional subspace $H_{m-r},$ let $t=R_{H_{m-r}}.$
We have $m-2r\leq t\leq m-r.$
By Proposition 2, we can choose such an $(r-1)$-dimensional subspace
$ G_{r-1}=\langle \beta_{1},\beta_{2},\ldots,\beta_{r-1}\rangle $
that $ G_{r-1}\subset G^{\bot}_{r-1}.$ So the rank of $f$
over $G^{\bot}_{r-1}$ is $m-2r+2.$ By Proposition 1,
$|D_{a}\bigcap G^{\bot}_{r-1}|=p^{m-r}\pm p^{\frac{m-1}{2}}>0$
with $ a\in F^{*}_{p}.$
So we can choose an nonzero element $\beta_{r}\in D_{a}\bigcap G^{\bot}_{r-1}$
and get an $r$-dimensional subspace
$ H_{r}=\langle \beta_{1},\beta_{2},\ldots,\beta_{r-1},\beta_{r}\rangle.$
It is easy to see that the rank of $f$ over $H_{r}$ is $1$
and $ H_{r}\bigcap H^{\bot}_{r}=G_{r-1}.$
So the rank of $f$ over $H^{\bot}_{r}$ is $m-2r+1.$
We construct an $(m-r)$-dimensional subspace $H_{m-r}=H^{\bot}_{r}$ and set
$$
H_{m-r}=
\langle \alpha_{1},\alpha_{2},\ldots,\alpha_{m-2r+1},\beta_{1},\beta_{2},\ldots,\beta_{r-1}\rangle.
$$
Let $H_{m-2r+1}=\langle \alpha_{1},\alpha_{2},\ldots,\alpha_{m-2r+1}\rangle$ and set
$$
H^{\bot}_{m-2r+1}=\langle \beta_{1},\beta_{2},\ldots,\beta_{r-1},\beta_{r},\beta_{r+1},\ldots,\beta_{2r-1}\rangle.
$$
Because $R_{H_{m-r}}=m-2r+1,$ we get $H^{\bot}_{m-2r+1}\bigcap H_{m-2r+1}=\{0\}$ and
$$
\Delta_{m}=\Delta(\alpha_{1},\alpha_{2},\ldots,\alpha_{m-2r+1},
\beta_{1},\beta_{2},\ldots,\beta_{r-1},\beta_{r},\beta_{r+1},\ldots,\beta_{2r-1})
$$
$$
=\Delta_{H_{m-r}}\cdot\Delta(\beta_{1},\beta_{2},\ldots,\beta_{r-1},\beta_{r},\beta_{r+1},\ldots,\beta_{2r-1})
$$
$$
=(-1)^{r}\Delta_{H_{m-r}}a\det(M^{2}).
$$
Here $M$ is the square matrix $M=(F(\beta_{i},\beta_{j}))_{1\leq i\leq r-1,r+1\leq j\leq 2r-1}.$
According to the selection of $\beta_{r},$ we know that $ \overline{\eta}(\Delta_{H_{m-r}})$ may take $1$ or $-1.$
Hence for an $(m-r)$-dimensional subspace $H_{m-r}$ with $R_{H_{m-r}}= m-2r+1, $ by Proposition 1,
we have $|\overline{D}_{0}\bigcap H_{m-r}|=p^{m-r-1}\pm(p-1)p^{\frac{m-3}{2}}.$
Applying Proposition 1 once again, we have that when $2\leq r< \frac{m}{2},$
$\max\{|\overline{D}_{0} \bigcap H|: H \in [F_{p^{m}},m-r]_{p}\}=p^{m-r-1}+(p-1)p^{\frac{m-3}{2}}.$
Using Lemma 1, we can obtain the results in the theorem. The proof is finished.

\par \vskip 0.2 cm
\section{Concluding Remarks}
Quadratic forms are ones of the well-known polynomials in the theory of algebra.
In the paper, we restrict quadratic forms over finite fields of odd characteristic
on their subspaces and prove the existence of some special
subspaces and their dual spaces related to non-degenerate quadratic forms.
Then we determine completely the weight hierarchy of a class of linear codes 
from non-degenerate quadratic forms. We hope that our results may help to 
research the generalized Hamming weight of other linear codes.

\par  \vskip 0.5 cm


\begin{thebibliography}{}
%
%
\bibitem{1BL14}
M. Bras-Amor¨®s, K. Lee, and A. Vico-Oton, New lower bounds on the
generalized Hamming weights of AG codes, IEEE Trans. Inf. Theory,
60(10), 5930-5937(2014).
\bibitem{2BM00}
A. I. Barbero and C. Munuera, The weight hierarchy of
Hermitian codes, SIAM J. Discrete Math., 13(1), 79-104(2000).

\bibitem{3CC97}
J. Cheng and C.-C. Chao, On generalized Hamming weights of binary
primitive BCH codes with minimum distance one less than a power of
two, IEEE Trans. Inf. Theory, 43(1), 294-298(1997).

\bibitem{5DJ15}
C. Ding, Linear codes from some 2-designs, IEEE Trans. Inf. Theory, 61(6), 3265-3275(2015).

\bibitem{6DD14}
K. Ding,  C. Ding, Bianry linear codes with three weights, IEEE Communication Letters,
18(11), 1879-1882(2014).

\bibitem{7DF14}
M. Delgado, J. I. Farr¨¢n, P. A. Garc¨ªa-S¨¢nchez, and D. Llena, On
the weight hierarchy of codes coming from semigroups with two
generators, IEEE Trans. Inf. Theory, 60(1), 282-295(2014).

\bibitem{8DW17}
Xiaoni Du and Yunqi Wan, Linear codes from quadratic forms, Applicable Algebra in
Engineering, Communication and Computing, doi:10.1007/s00200-017-0319-x(2017).

\bibitem{9DL16}
C. Ding, C. Li, N.Li, Z. Zhou, Three-weight cyclic codes and their weight distributions. Discret.
Math., 339(2), 415-427 (2016).

\bibitem{11HP98}
P. Heijnen and R. Pellikaan, Generalized Hamming weights of
q-ary Reed¨CMuller codes, IEEE Trans. Inf. Theory, 44(1), 181-196(1998).

\bibitem{12HP03}
W. C. Huffman and V. Pless,  Fundamentals of error-correcting codes,
Cambridge University Press, Cambridge(2003).

\bibitem{13JF17}
G. Jian, R. Feng and H. Wu, Generalized Hamming weights of
three classes of linear codes, Finite Fields and Their Applications, 45, 341-354(2017).

\bibitem{14JL97}
H. Janwa and A. K. Lal, On the generalized Hamming weights of cyclic
codes, IEEE Trans. Inf. Theory, 43(1), 299-308(1997).

\bibitem{15KJ78}
T. Kl$\phi$ve, The weight distribution of
linear codes over $GF(q^{l})$ having generator matrix over $GF(q>)$, Discrete Math., 23(2), 159-168(1978).

\bibitem{16LN97}
 R. Lidl, H. Niederreiter,  Finite fields, Cambridge University Press, New York(1997).

\bibitem{17LY14}
C. Li, Q. Yue, and F. Li, Hamming weights of the duals of cyclic codes
with two zeros, IEEE Trans. Inf. Theory, 60(7), 3895-3902(2014).

\bibitem{18LF17}
F. Li, A class of cyclotomic linear codes and their generalized Hamming weights, arXiv:1708.04415(2017).

\bibitem{19TV95}
M. A. Tsfasman, S. G. Vladut, Geometric approach to higher weights, IEEE Trans. Inf.
Theory, 41(6), 1564-1588(1995).

\bibitem{20WJ91}
V.K.Wei,  Generalized Hamming weights for linear codes, IEEE Trans. Inf. Theory, 37(5), 1412-1418(1991).

\bibitem{21XL16}
M. Xiong, S. Li, and G. Ge. The weight hierarchy of some reducible cyclic
codes. IEEE Trans. Inf. Theory, 62(7), 4071-4080(2016).

\bibitem{22YL15}
M. Yang, J. Li, K. Feng and D. Lin, Generalized Hamming weights of
irreducible cyclic codes, IEEE Trans. Inf. Theory, 61(9), 4905-4913(2015).

\bibitem{23YY17}
S.Yang, Z.Yao, Complete weight enumerators of a family of three-weight linear codes. Des. Codes
Cryptogr. 82(3), 663-674(2017).

\bibitem{24ZL16}
Z. Zhou, N. Li, C. Fan and T. Helleseth, Linear codes with two or three weights from quadratic bent functions, Des. Codes Cryptogr., 81(2), 283-295(2016).

\bibitem{25DD15}
K. Ding,  C. Ding, A class of two-weight and three-weight codes
and their applications in secret sharing, IEEE Trans. Inf. Theory,
61(11), 5835-5842(2015).

\bibitem{26TX17}
C. Tang, C. Xiang and K. Feng,  Linear codes with few weights from inhomogeneous
quadratic functions, Des. Codes Cryptogr., 83(3), 691--714(2017).

\bibitem{27WZ94}
Z. Wan, The weight hierarchies of the projective codes
from nondegenerate quadrics, Des. Codes Cryptogr., 4(4), 283-300(1994).
\end{thebibliography}


\end{document}